\documentclass[12pt]{article}
\usepackage{times}
\usepackage{geometry}
\usepackage[utf8]{inputenc}
\usepackage{enumitem,amssymb}
\usepackage{ragged2e}
\newlist{thematic}{itemize}{8}
\setlist[thematic]{label=$\square$}
\usepackage{pifont}
\usepackage{amssymb, amsmath} 
\usepackage{fancyhdr}
\usepackage[numbers,sort, compress]{natbib}
\usepackage{graphics,graphicx}
\usepackage{color}
\usepackage[table]{xcolor}
\usepackage{colortbl} 
\usepackage{tabularx}
\usepackage[para]{footmisc}
\usepackage{caption}
\usepackage{paralist} 
\usepackage[colorlinks=true, citecolor=blue, linkcolor=blue, urlcolor=blue]{hyperref}
\usepackage{wrapfig}
\newcommand{\spbefore}{-0.45cm}
\newcommand{\spafter}{-0.25cm}
\def\apj{\rm ApJ}
\def\apjl{\rm ApJL}
\def\apjs{\rm ApJS}
\def\aj{\rm AJ}
\def\mnras{\rm MNRAS}
\def\araa{\rm ARAA}
\def\nat{\rm Nature}

\def\pasp{\rm PASP}
\def\aap{\rm A\&A}
\def\prd{\rm PRD}

\begin{document}
\begin{flushleft}

\huge
Astro2020 Science White Paper \linebreak

Multi-Physics of AGN Jets in the Multi-Messenger Era
\linebreak
\normalsize

\noindent \textbf{Thematic Areas:}
 $\text{\rlap{$\checkmark$}}\square$          Multi-Messenger Astronomy and Astrophysics \hspace*{65pt} \linebreak
  
\textbf{Principal Author:}
Name: Bindu Rani
 \linebreak						
Institution:  NASA Goddard Space Flight Center, Greenbelt, MD, USA 
 \linebreak
Email: bindu.rani@nasa.gov; Phone: +1 301.286.2531
 \linebreak
 \textbf{Lead authors:}  M.~Petropoulou (Princeton University, USA), H.~Zhang (Purdue University, USA), F.~D'Ammando (INAF, Italy), and J.~Finke (NRL, USA) 
 \linebreak
\textbf{Co-authors:} {\small M. Baring (Rice University, USA), M.~B{\"o}ttcher (North-West University, South Africa), S.~Dimitrakoudis (University of Alberta, Canada), Z.~Gan (CCA, USA), D.~Giannios (Purdue University, USA),  D. H. Hartmann (Clemson University, USA), T.~P.~Krichbaum (MPIfR, Germany), A.\ P.\ Marscher (Boston University, USA), A.~Mastichiadis (University of Athens, Greece), K.~Nalewajko (Nicolaus Copernicus Astronomical Center, Poland), R.~Ojha (UMBC/NASA GSFC, USA), D.~Paneque (MPP, Germany), C.~Shrader (NASA GSFC, USA), L.~Sironi (Columbia University, USA), A.~Tchekhovskoy (Northwestern University, USA), D.~J.~ Thompson (NASA GSFC, USA), N.~Vlahakis (University of Athens, Greece), T. M. Venters (NASA GSFC, USA)}
%
\end{flushleft}


\begin{figure}[h] 
    \centering
    \includegraphics[width=0.85\textwidth, trim=0 0 -60 0]{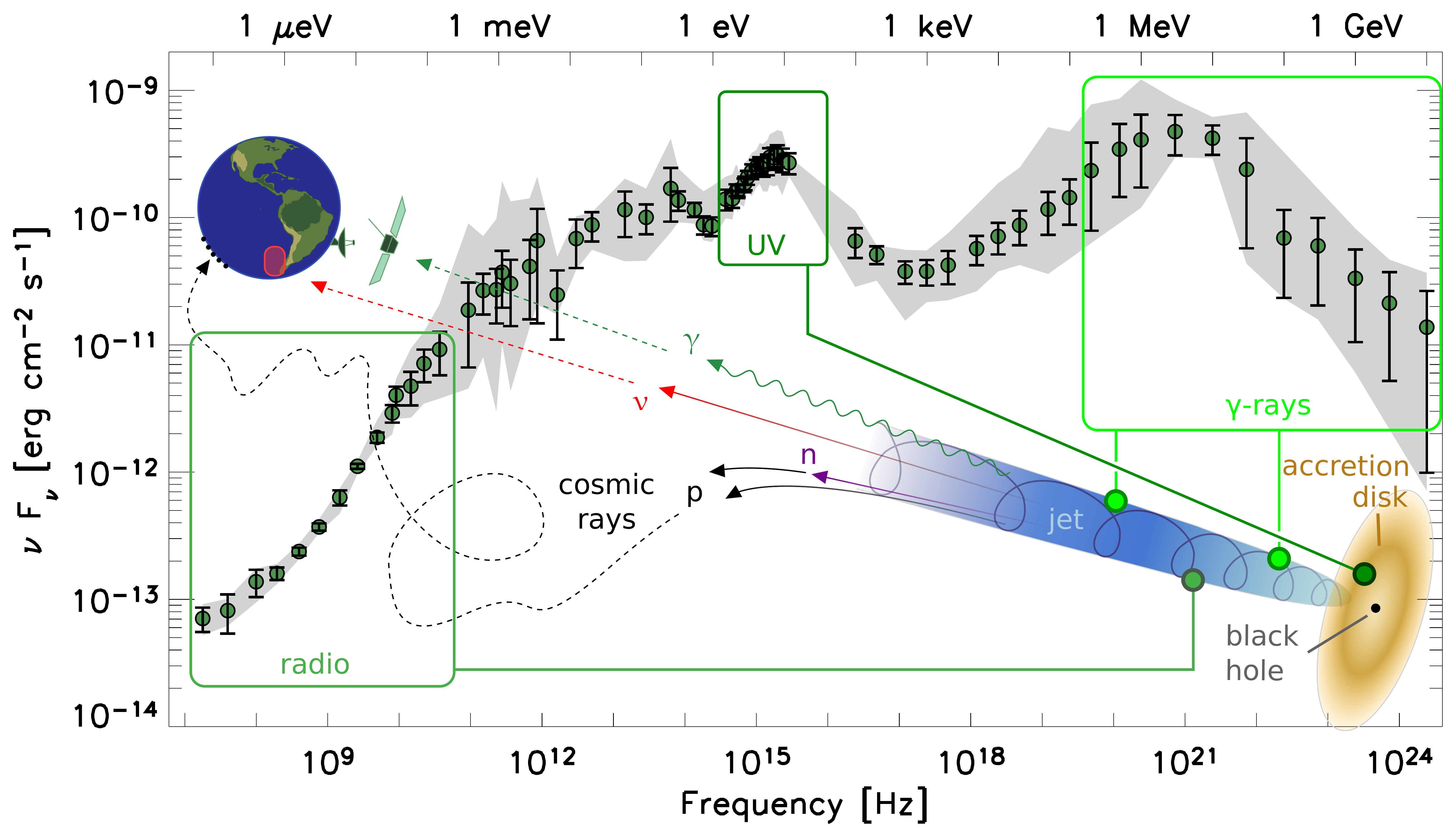}
    {\caption{{AGN jets, powered by accretion onto a central supermassive black hole, are the most powerful and long-lived particle accelerators in the Universe. Non-thermal processes operating in jets are responsible for multi-messenger emissions, such as broadband electromagnetic radiation and high-energy neutrinos. Background spectral energy distribution is adapted from \citep{Soldi2008}.} }
    \label{fig:sketch}
    }
\end{figure}
\setlength{\parindent}{0.25in}

\pagebreak

\vspace{\spbefore}
\section{Scientific context}
\vspace{\spafter}
Active galactic nuclei (AGN) with relativistic jets, powered by gas accretion onto their central supermassive 
black hole (SMBH), are  \emph{the most powerful and long-lived particle accelerators in the Universe}. They are unique 
laboratories for studying the physics of matter and elementary particles in extreme conditions (e.g. in strong gravity, 
magnetic fields, low matter density, and high energy density plasmas moving at relativistic speeds) that cannot be realized 
on Earth. AGN jets exhibit highly variable non-thermal emission across the entire electromagnetic (EM) spectrum, from radio 
up to TeV $\gamma$-rays \citep{aharonian07,Albert07,ackermann16}. Blazars, a subclass of AGN having their jets pointed very 
close to our line of sight, are the most numerous extragalactic $\gamma$-ray sources. For a long time since the discovery of AGN, 
photons were the only way to probe the underlying physical processes. The recent discovery of a very high energy neutrino, 
IceCube-170922A, coincident with a flaring blazar, TXS 0506+056 \citep{ice18a}, provides the first evidence that AGN jets 
are \emph{ multi-messenger} sources; they are capable of accelerating hadrons to very high energies, while producing non-thermal 
EM radiation and high-energy neutrinos.
This new era of multi-messenger astronomy, which will mature in the next decade, offers us the unprecedented opportunity to combine 
more than one messenger to solve some long-standing puzzles of AGN jet physics: {\it How do jets dissipate their energy to 
accelerate particles? What is the jet total kinetic power? Where and how do jets produce high-energy emission and neutrinos?  
What physical mechanisms drive the particle acceleration? } 

\vspace{\spbefore}
 \section{Key science questions of AGN jet physics}   
\vspace{\spafter}
\subsection{What are the dissipation and particle acceleration processes?}\label{sec:acceleration}
The detection of AGN at high energies as well as the non-thermal nature of their emission provide strong evidence for 
particle acceleration in jets. The inferred radiative efficiency of AGN jets is typically 10\% \citep[or higher in 
pair-loaded jets;][]{celotti08, ghisellini14}, suggesting an efficient conversion of the jet energy into radiation. 
Neither the acceleration processes nor the radiation mechanisms can be fully understood independent of the jet dynamics 
(i.e., launching, acceleration, propagation,  collimation, and dissipation).
There is strong theoretical motivation that jets are launched as magnetically dominated flows at their 
base  \citep[e.g.,][]{blandford_znajek77,lyubarsky10, tchekhovskoy11}. The strong magnetic fields threading a 
rotating compact object or the associated accretion disk can convert the rotational energy of the central engine into the 
power of the outflow \citep{blandford_znajek77, blandford_payne82}. Magnetohydronamic (MHD) jet models find that part of the 
jet magnetic energy is used to accelerate the bulk flow to relativistic speeds, over a wide range of scales, i.e., $\sim$sub-parsecs  
to parsecs (pc) \citep[e.g.][]{vlahakis_koenigl04, li06,lyubarsky09, komissarov09}. However, the nature of the flow (i.e., kinetically 
versus magnetically dominated) at the scales where the jet dissipates its energy to accelerate the radiating particles is largely 
unknown (see also \S\ref{sec:sites}).

The particle acceleration mechanism behind the jet emission strongly depends on the energy composition and evolution of the flow. Kinetically 
dominated flows can dissipate their energy and efficiently accelerate particles at relativistic \emph{shocks} via the diffusive shock 
acceleration (DSA) mechanism \citep{heavens88,kirk00,achterberg01,spitkovsky08,summerlin12}. The predictions of DSA theory have been 
widely applied to jet observations \citep{marscher85,spada01,Joshi07,Graff08,larionov13,Chen14,zhang16a,boettcher19}. However, the 
acceleration efficiency can be greatly reduced in strongly magnetized flows \citep{kennel84, kirk89, gallant92, ss09, ss11, summerlin12,sironi15}. 
In the regime where magnetic fields are dynamically important, {\it magnetic reconnection} can efficiently tap magnetic energy to accelerate 
particles \citep{romanova92, giannios09, sironietal15, petroetal16, zhang18, nalewajko18, christie19}. This non-ideal MHD process can transfer 
magnetic energy to particles when magnetic field lines of opposite polarity annihilate. Numerical studies in both pair and electron-proton 
plasmas have shown the formation of an extended power-law in the particle energy distribution that are also consistent with 
observations \citep{melzani14,guo14, ss14, Guo16, Deng16, werner16}. These particles can be accelerated at reconnection X-points, magnetic 
islands, and secondary current sheets in-between merging plasmoids \citep[e.g.][]{ss14, nalewajko15, ps18}. It is also likely that the flow 
is not laminar but {\it turbulent} at the dissipation sites. Theoretical studies suggested that shocks in a turbulent environment can accelerate 
particles and lead to stochastic radiation and polarization signatures \citep{larionov13,marscher14,baring17,tavecchio18}. Recent particle-in-cell 
(PIC) simulations of turbulence-driven reconnection also demonstrated the generation of non-thermal particles \cite{comisso18, zhdankin18}. These 
are mainly accelerated by stochastic scattering on the large-scale magnetic fluctuations (i.e., via second-order Fermi process) in contrast to 
magnetic reconnection in laminar flows.

{But which processes dominate the jet energy dissipation and particle acceleration? How turbulent is the jet at the dissipation sites? And how can 
we disentangle these processes?} 
\emph{Next-generation multi-physics simulations} that self-consistently connect fluid dynamics, particle acceleration,  and radiation,
will study energy dissipation and particle acceleration mechanisms under different jet physical conditions and to deliver temporal and 
spatial information about radiation and polarization \citep[for preliminary efforts see][]{christie19,tavecchio18,zhang18}.
In combination to simultaneous \emph{high-resolution radio interferometric observations}  at millimeter wavelengths (VLBI) 
and \emph{polarization measurements} at optical wavelengths \citep{king14, pavlidou14} and higher energies 
(by future X-ray/$\gamma$-ray polarization missions), we will be able to settle these questions in the next decade.

\vspace{\spbefore}
\subsection{What are the high-energy radiation mechanisms?}\label{sec:radiation}
The observed broadband radiation from an AGN usually follows a double humped structure \citep[Fig.~\ref{fig:sketch}][]{ulrichetal97,fossatietal98}. 
In sources with luminous accretion disks, like flat spectrum radio quasars (FSRQs), there is also some thermal contribution to the blue/ultraviolet 
part of the spectrum. 
It is well established that the low-energy hump (radio to X-rays) is explained as synchrotron radiation from  relativistic electrons in the 
jet, but their acceleration mechanism is still debated. The emission processes as well as the type of radiating particles (i.e., electrons 
or protons) responsible for the high-energy (HE) hump  (hard X-rays to $\gamma$-rays) are also unresolved. \emph{Leptonic} scenarios have been 
put forward to explain the HE hump as a result of inverse Compton (IC) scattering processes by relativistic electrons on a photon field. The 
seed photons for the scattering can be external to the jet (from the accretion disk, broad line region (BLR) or dusty torus) and/or the low-energy 
synchrotron hump \citep[e.g.][]{maraschietal92, dermeretal92, sikoraetal94,bloommarscher96,ghisellinimadau96,boettcherdermer98}. The same process 
that accelerates electrons to relativistic energies can also act upon protons that are present in the jet. In fact, protons may reach much higher 
energies than electrons, as they are not strongly affected by radiative losses. The presence of a relativistic hadronic component in the jet is the 
cornerstone of \emph{leptohadronic} scenarios that attribute the HE AGN emission to interactions involving relativistic protons: proton-induced EM 
cascades~\citep[e.g.,][]{mannheim93,muecke03, mastetal13}, neutral pion decays \citep{sahu13,cao14}, or proton synchrotron 
radiation~\citep{aharonian00,mueckeprotheroe01,muecke03}. 

So far, both classes of models have been successful in describing the broadband jet 
emission \cite[e.g.][]{celotti08, boettcher13, petro14}, with leptohadronic scenarios typically requiring high 
jet powers that strain theoretical jet launching models \citep[e.g.][]{zdziarski15,petroetal15, petrodimi15}. Such high-power 
jets may also have a larger impact on their environments compared to low-power jets. 
There is a large degeneracy in the theoretical description of the data, which allows any 
given broadband SED to be described within many variations of a given theoretical 
model \citep{2017A&A...603A..31A}, or even within very distinct  theoretical 
frameworks, like a purely leptonic and a purely hadronic \citep{2011ApJ...736..131A}. 
\emph{Model predictions that go beyond the time-averaged spectral properties are necessary to distinguish 
between the two main scenarios of HE emission in AGN.} Pure leptonic AGN models do not predict any neutrino emission, 
since the latter can be produced only through interactions of relativistic protons with matter \citep[inelastic pp collisions, e.g.][]{kelner06} or radiation \citep[photohadronic interactions;][]{muecke00, kelner08}. Although a secure (i.e., high significance) association of HE neutrinos with AGN will be the ``smoking gun'' of proton acceleration in AGN jets \citep[e.g.][]{petroetal15, petromast15, keivani18}, one can take advantage of the accompanying EM signals. For example, the synchrotron radiation of pairs produced by photohadronic interactions may emerge as a third photon component in the $\sim$40 keV -- 40 MeV energy range, a  poorly
explored regime of the EM spectrum \citep{petromast15}. An additional component from proton-induced cascades may emerge at energies $\gtrsim 0.1- 1$~TeV \citep[e.g.][]{sol13, petrodimi15}. Both components may introduce unique temporal variability signatures \citep[e.g.][]{mastetal13, diltz15, petroetal17}. In addition, HE  polarization can also pinpoint the leptonic and hadronic radiation mechanisms \citep{Krawczynski12, zhang13, zhang16,paliya18}. \emph{These theoretical predictions can be confronted with future multi-messenger experiments to unveil the radiation processes in AGN jets.}

\vspace{\spbefore}
\subsection{Where are the high-energy dissipation sites located?}\label{sec:sites}
The location of HE emission in AGN jets has been a long standing question. It is well known that the radio emission in AGN jets must 
be produced in a large enough region to avoid synchrotron self-absorption \citep[e.g.][]{marscher80, marscher85, potter12}. This requirement  
places the radio-emitting region at pc-scale distances from the central SMBH. Rapid variability at optical and $\gamma$-ray 
energies \citep[e.g.,][]{aleksic11}, on the other hand, indicates a much more compact emitting region than that of the radio emission. 
Some leptonic emission models suggest that the rapidly variable emission region is located within the BLR ($\sim 0.1~\rm{pc}$ from the SMBH) 
and takes up the entire jet cross-section. However, evidence has been building for some time that a significant amount of emission may occur 
beyond the BLR, at pc distances from the SMBH \citep[e.g.,][]{costamante2018, meyer19}. Observations with the Very Long Baseline Array (VLBA) 
indicate a connection between the formation of a new jet feature on the pc scale and a strong $\gamma$-ray flare in blazars \citep{marscher08, orienti13,jorstad17}.  If $\gamma$ rays with energies $\gg 10$~GeV are produced within the BLR, they will interact with BLR photons and produce absorption features at $\gtrsim10$\ GeV, which are hardly seen in 6 years of LAT data \citep{costamante2018} and at very high energies (VHE, $\gtrsim100$\ GeV) \citep[e.g.,][]{albert2008_3c279,aleksic11}. HE emission on  pc scales is further supported by the localization of $\gamma$-ray flares with gravitational lensing \citep{barnacka2015,barnacka2016} and Gaia optical position offsets from the VLBI radio core position \citep{plavin2018}.

Various physical processes predict the formation of very compact dissipation regions at various distances from the SMBH, including relativistic turbulence, magnetic reconnection, kink instabilities, and recollimation shocks \citep[e.g.][]{marscher14,giannios13,nalewajko12}. Which one dominates at different scales is directly related to the jet propagation through the surrounding medium \citep{Guan14,Barniol18}. With the introduction of GPU-based computing, large-scale MHD simulations of jets from the launching to the dissipation site become now possible \citep{Liska18,Schneider17,Kestener17}. When combined with particle acceleration and radiative transfer, \emph{these studies will unveil the physics of the HE emission region in the next decade. Spectral and temporal information from 0.1 GeV to multi-TeV energies and high-resolution radio interferometric images at short radio wavelengths will be crucial to test these theoretical predictions.}

\vspace{\spbefore}
\subsection{Is the $\gamma$-ray emission related to the jet structure?}
Contrary to the other blazar sub-populations, many TeV BL Lacs and radio galaxies (detected above 0.5~TeV) show neither substantial superluminal jet components nor high variability in radio, indicating that no strong Doppler beaming takes place in their jet, at least on scales probed by radio observations. These results are in tension with the high Doppler factors usually required in the modeling of the high-energy emission, commonly known as the TeV Doppler factor crisis. A possible explanation is that the jets of TeV BL~Lacs are not homogeneous but stratified, showing either a radial \citep{georganopoulos03} or a transverse velocity gradient \citep{ghisellini05}. 
Theoretical arguments and numerical simulations suggest that jets are not uniform outflows, but are characterized by a transverse velocity structure composed of a fast central part, the spine, surrounded by a slower layer \citep{ferrari98}. The regions with different speeds would interact through their radiation fields, relativistically boosted in the different frames. Such interaction leads to the enhancement of the IC emission of the two zones. Another consequence of the radiative coupling is the progressive deceleration of the spine, an effect that may explain the modest speeds found in most of the TeV BL Lacs through VLBI observations. Moreover, since the layer is expected to have lower values of the bulk Lorentz factor with respect to the spine, its beamed emission is lower, and can be detected even when the jet is misaligned with respect to us, as in the case of radio galaxies.  Strong support to the existence of a stratified jet structure in BL Lacs  and FR I radio galaxies\footnote{Radio galaxies are classified based on whether they are brighter near the lobes (FR II) or at the center \citep[FR I;][]{fanaroff74}. FR I and FR II are also lower and higher power sources, respectively \citep[e.g.,][]{fanaroff74,ledlow96}.}, i.e. their misaligned parent population \citep{urry95}, comes from the observation of limb-brightened structures in the TeV BL Lac Mrk 501 \citep{giroletti04}, the TeV radio galaxies M87 \citep{hada13}, 3C 84 \citep{nagai14}, and in the misaligned blazar PKS 0521-36 \citep{dammando15}. \emph{It is not clear whether the most powerful jets, i.e. those of FSRQs and FR II radio galaxies, have similar structures.} Differences in the jet structure between powerful and weak sources could be related to different environments enshrouding the jet and/or to intrinsic jet properties, causing the weak jets to be more prone to instabilities \citep{tchekhovskoy16}. Interestingly, there is observational evidence for differences in the environments and accretion modes of FR I and FR II galaxies \citep{best12}.  

An alternative explanation is the ``jets in a jet'' model  \citep{giannios09, petroetal16, christie19}, according to which the rapid, highly beamed emission comes from  ``mini-jets'' with much higher Lorentz factors than the one of the surrounding bulk flow that is responsible for the radio emission. There is a well-founded theory behind the formation of these mini-jets, which is directly related to jet dissipation mechanisms (\S\ \ref{sec:acceleration}). A single reconnection layer formed in the jet can be envisioned as a mini-jet that can be oriented in almost any direction relative to the larger jet flow, while in the spine-layer model, the high-Doppler factor component would have to have the same alignment as the slower jet flow. The angle to the line of sight of some TeV BL Lacs is large, favoring the  reconnection-driven mini-jets model \citep{finke2018}. Still, a better understanding of the statistics and polarization properties of 
mini-jets and of their coupling with the slower flow is needed.
\emph{Direct comparison of observations with theoretical models of jet dissipation and a study of the jet angle to the line of sight with VLBI will be crucial in understanding high-energy dissipation conditions. }  
Millimeter-VLBI observations will  image the central AGN and inner jet regions
on spatial scales $\leq$100 gravitational radii.

\vspace{\spbefore}
\section{Key advances in instrumentation, theory and simulation}
\vspace{\spafter}

\begin{figure}[t]
    \centering
        {\caption*{\small Table 1: Key scientific questions and future developments in instrumentation, theory \& simulation.}}
    \includegraphics[width=\textwidth]{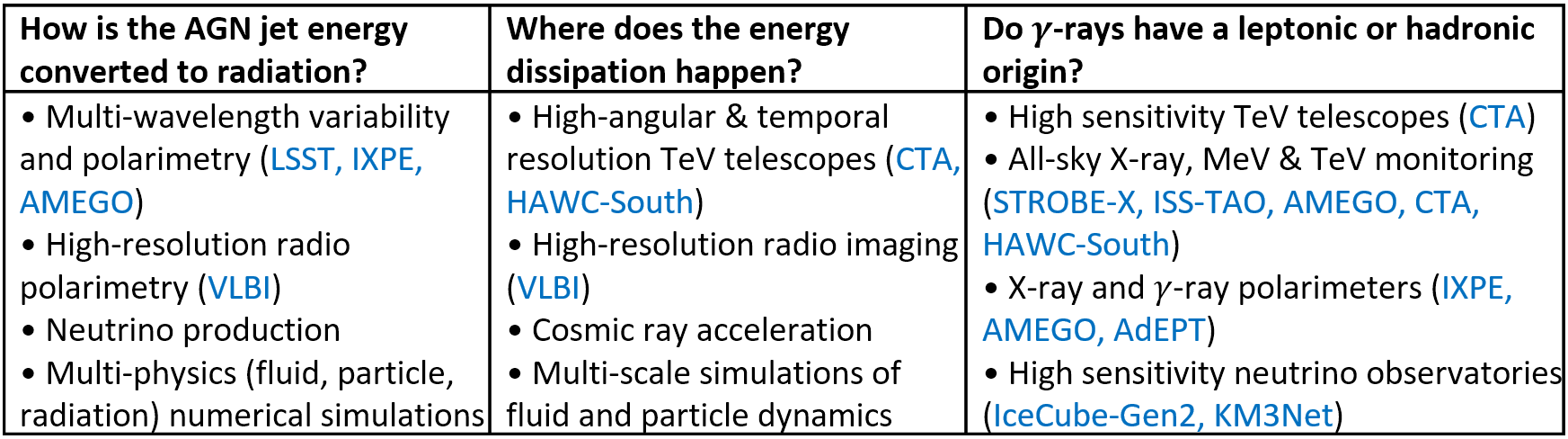}
    \vspace{-1.cm}
  \label{fig:table}
\end{figure}
More associations of HE neutrinos with individual AGN are necessary for making a strong claim for their physical connection. The next-generation neutrino observatory with $\sim5$ times the point-source sensitivity of IceCube ({\bf IceCube-Gen2\footnote{\url{https://icecube.wisc.edu/science/beyond}}} \cite{icecube_gen2}) will lead to order-of-magnitude increase of the source detection rates \citep{ahlers14}.
The Cherenkov Telescope Array ({\bf CTA\footnote{\url{https://www.cta-observatory.org/}} }\cite{cta2019})
will play a major role in future searches for the detection of spectral features resulting from hadronic processes at TeV energies (see \S\ref{sec:radiation}), which require high sensitivity and high spectral resolution. If the GeV AGN emission is produced by hadronic processes, then a bright component between the low- and high-energy humps of the SED is also expected (\S\ref{sec:radiation}). MeV observations of AGN (in both quiescent and flaring states) with future missions like All-sky Medium Energy Gamma-ray Observatory ({\bf AMEGO\footnote{\url{https://asd.gsfc.nasa.gov/amego/}}}, \cite{amego2019}), acting complementary to neutrino searches, could probe the HE hadronic component of AGN jets by detecting or setting upper limits to the predicted MeV emission. Sensitive all-sky X-ray monitors, like {\bf ISS-TAO\footnote{\url{https://asd.gsfc.nasa.gov/isstao/}}} and {\bf STROBE-X\footnote{\url{https://gammaray.nsstc.nasa.gov/Strobe-X/}}}  \citep{strobeX}, will be ideal for searches of {EM counterparts to HE neutrino events} and for delivering uninterrupted X-ray light curves of bright AGN. HE polarimetric missions, like {\bf IXPE\footnote{\url{https://ixpe.msfc.nasa.gov/}}} \citep{ixpe}, {\bf AMEGO}, and  Advanced Energetic Pair Telescope ({\bf AdEPT} \cite{adept}),  will provide polarization observations of bright AGN at hard X-rays and MeV $\gamma$-rays energies, which will disentangle the competing HE emission models (leptonic versus hadronic).  {\bf We advocate the support to future instruments with large effective areas, excellent timing resolution, and wide fields of view  that will be essential for advancing our understanding of jet physics in the next decade.}

\smallskip

Current theoretical studies of AGN jets treat the fluid dynamics, particle acceleration, and radiation processes separately, due to the huge differences in physical scales. CPU-based codes, for example, have been successful in simulating the 3D dynamics of jets and particle acceleration,  but without directly connecting the fluid dynamical scales with particle kinetic scales \citep[e.g.,][]{li06,tchekhovskoy11,ss14,guo14,zhang17, petro18}. However, the jet dynamics and interactions with the surrounding medium will determine the location of energy dissipation, which sets the physical conditions for particle acceleration and transport. These particles will radiate and interact with photons, which will in turn feedback onto the particle evolution. Current techniques for the radiative transfer calculations, which involve Monte Carlo tracing and particle transport equation solvers, often neglect the acceleration processes as well as spatial particle advection and diffusion \citep[e.g.,][]{dmpr12, boettcher13,Chen14,Zhang14,petroetal15, cerruti15,MacDonald18}.
\emph{The complexity of these problems calls for support to the development of multi-physics, multi-scale numerical simulations, and high-performance computing.}
New developments in GPU-based computing \citep{Liska18,Schneider17,Kestener17} will allow the simultaneous treatment of micro- and macro-physics in jet simulations. With the addition of radiative feedback physics (e.g., absorption, scattering, hadronic interactions, and polarization) on top of the multi-scale jet simulations, we will be able to critically test our theories for AGN jet physics against the multi-messenger observations in the next decade.

\bigskip 

\section*{Acknowledgement}
We acknowledge the support of NASA's Physics of Cosmos Multimessenger Astrophysics Science Analysis Group in  the organization of the white paper. 


\end{document}